\documentclass[11pt,english]{article}
\usepackage{babel} % DEUTSCHE SONDERZEICHEN
\usepackage[ansinew]{inputenc} % UMLAUTE
\usepackage{pdflscape}
\usepackage{setspace} % ZEILENABST\HAT{\THETA_I}NDE US W
\usepackage{amsmath} % MATHE
\usepackage{comment} 
\usepackage{amssymb} % MATHE
\usepackage[table]{xcolor}
\usepackage{bbm}
\usepackage{booktabs}
\usepackage{amsfonts} % MATHE
\usepackage{titling}
\usepackage[small]{caption}
\usepackage{geometry}
\usepackage{multirow,array}
\usepackage{amsthm}
\usepackage{epic,eepic}
\usepackage{hyperref}
\usepackage{multirow}
\usepackage{bbm}
\usepackage{dcolumn}
\usepackage{adjustbox}
\usepackage{breakurl}
\usepackage{subfigure}
\usepackage{xurl}
\usepackage{rotating}
\usepackage{array}
\usepackage{natbib}

\newcommand{\ot}{\frac{1}{2}}
\newcommand{\ps}{ pure strategy }
\newcommand{\hp}{\hat{\pi}}
\newcommand{\hm}{\tilde{\mu}}
\newcommand{\tp}{\tilde{\pi}}
\newcommand{\kps}{(s,\mathbf{p},\hm)}
\newcommand{\kpss}{(s,\mathbf{p},\hm^*)}

\usepackage{breakcites}
\usepackage{graphicx}
\usepackage[UKenglish]{datetime}
\usepackage{color}
\newtheorem{prop}{Proposition}

\newtheorem{assume}{Assumption}
\newtheorem{lemma}{Lemma}
\newtheorem{coroll}{Corollary}
\selectfont

\begin{document}

\setlength{\droptitle}{-2cm}

\title{\textbf{Motivated Reasoning and the Political Economy of Climate Change Inaction}\thanks{I would like to thank Benjamin Blumenthal, Agustin Casas, Boris Ginzburg, Johannes Schneider,  Dana Sisak, and Arne  Weiss as well as the participants of the 2024 Microecononomics  Workshop in L\"{u}neburg, the 2024 EPCS congress in Vienna, and the 2025 EPSA conference in Madrid for valuable comments and suggestions. The hospitality of DICE at University of D\"{u}sseldorf, where parts of this paper were written, as well as the financial support of the Agencia Estal de Investigacion (Spain) through grants PID2022-
141823NA-I00, MICIN/ AEI/10.13039/501100011033, and  CEX2021-001181-M as well as of the Comunidad de Madrid through grant EPUC3M11 (V PRICIT)  is gratefully acknowledged.}}
\author{Philipp Denter\thanks{Universidad Carlos III de Madrid, Department of Economics, Calle de Madrid 126, 29803 Getafe, Spain. E-Mail: \href{mailto:pdenter@eco.uc3m.es}{pdenter@eco.uc3m.es}.} }

\maketitle

\begin{abstract}
We study how motivated reasoning affects the provision of climate policy in an electoral competition framework. Voters experience anticipatory disutility when future outcomes appear grim and may therefore distort beliefs in response to adverse information. We develop a game-theoretic model in which voters and politicians receive signals about the severity of climate change.
When the anticipated welfare losses from severe climate change are sufficiently large, voters optimally ignore unfavorable information, inducing politicians to campaign on policies appropriate for mild climate change only. When welfare losses are moderate, the model admits a second, efficient equilibrium in which voters trust politicians to implement welfare-maximizing policies and vote informatively, thereby creating incentives for politicians to propose adequate climate policy. The model shows how motivated belief formation and voters' expectations about policy responsiveness jointly determine equilibrium selection between effective climate policy and persistent political inaction.
\end{abstract}

%%H12:  Public Economics-CRISIS MANAGEMNT


\newpage

\section{Introduction}
%
% JPUBE, AEJ: POLICY, AEJ: MICRO, JPE Micro, JEEM, Ecological Economics
% APSR, AJPS
%
%
%
%

\begin{flushright}
 \emph{``[\dots] information that increases
perceptions of the reality of climate\\
 change may feel so frightening that it
leads to denial and thus\\
a reduction in concern and support for action.''}

\vspace{3mm}

\cite{ClaytonManningKrygsmanSpeiser:2017}

\end{flushright}
%\vspace{5mm}
%
%\begin{flushright}
% \emph{``Yes, There Has Been Progress on Climate.\\ No, It's Not Nearly Enough.''}
%
%\vspace{3mm}
%
%The New York Times, 25 Oct 2021
%
%\end{flushright}

\vspace{5mm}

Climate change is one of the most pressing issues of our time, threatening the livelihoods of millions of people around the globe. Information about climate change and global warming has been available for more than a century, dating back at least to Svante Arrhenius' (1896) famous paper ``\emph{On the Influence of Carbonic Acid in the Air upon the Temperature of the Ground}.'' However, despite an ever-growing scientific consensus that increasing atmospheric CO2 concentrations will severely impact our planet, little to no action was taken to stop this process for most of the last few decades. Many reasons have been put forward to explain the inaction of political decision-makers, ranging from lobbying interests to widespread misinformation campaigns like those of ExxonMobil (e.g., \citealp{OreskesConway:2011merchants}, or \citealp{SupranRahmsdorfOreskes:2023}). In this paper, we discuss a different, simpler channel: the electoral incentives of politicians when voters may  hold motivated beliefs.
%

%Our modeling of awareness management is based on Bénabou and Tirole (2002)

The looming dire consequences of climate change may, if taken seriously, create stress and anxiety. In an effort to avoid these negative emotions, people may choose to hold motivated beliefs, ignoring information that suggests climate change is severe, while overreacting to information suggesting there is nothing to worry about. Psychologists and economists alike have long been aware of such information processing biases, as exemplified by Kunda's (1990) seminal paper or the recent survey by \cite{AmelioZimmermann:2023}, and multiple studies have provided evidence for their empirical relevance (for example,  \citealp{LewandowskyEtAl:2013},  \citealp{Thaler:2021,Thaler:2024}, and \citealp{LoisEtAl:2024}). 

\begin{figure}
    \centering
    \includegraphics[width=0.75\linewidth]{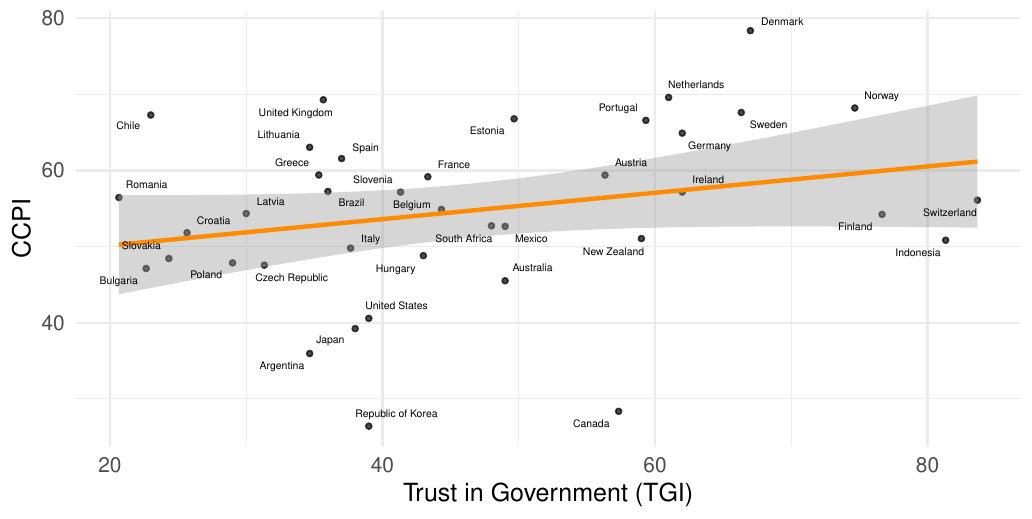}
    \caption{Scatterplot with fitted trendline and 95\% confidence interval showing countries' CCPI scores on the vertical axis and TGI scores on the horizontal axis.}
    \label{fig:scatter}
\end{figure}

Motivated reasoning alone, however, is unlikely to fully account for persistent political inaction on climate change. Recent empirical work has also emphasized the role of political trust in shaping climate policy. In particular, higher trust in political institutions has been associated with local politicians adopting more ambitious climate policies (\citealp{Bose:2023}). Figure~\ref{fig:scatter} shows that a similar positive relationship between political trust and climate policy stringency emerges in a cross-country comparison.\footnote{Data on climate policy performance come from the \emph{Climate Change Performance Index} (CCPI), published jointly by Germanwatch, the NewClimate Institute, and the Climate Action Network International. The CCPI ranges from 0 to 100, with higher values indicating better performance in climate policy. It aggregates indicators on greenhouse gas emissions (40\%), renewable energy (20\%), energy use (20\%), and other climate policies (20\%). Data on trust are taken from the OECD's \emph{Trust in Government Indicator} (TGI), which measures the share of a country's population reporting confidence in their national government and also ranges from 0 to 100.} These patterns point to an interaction between belief distortions and political trust in determining policy outcomes, a mechanism that the analysis below formalizes.

The aim of this paper is to construct a game-theoretic model that sheds light on how motivated reasoning affects politicians' incentives to propose adequate climate change policies within an electoral competition framework. In the model, there are two possible states of the world: severe climate change and mild climate change. The state determines which policy is optimal and establishes a baseline level of welfare. If climate change is severe, baseline welfare is lower than in the mild state, regardless of the policy chosen. Voters and politicians receive signals about the state of the world, and society is best served when the implemented policy matches the true state. As in \cite{brunnermeier2005optimal}, \cite{BenabouTirole:2016}, or \cite{Spiegler:2016}, voters derive anticipatory utility and experience anxiety or stress when the future appears grim. To cope with these negative emotions, they may adopt motivated beliefs, interpreting information about the future in a non-Bayesian manner to thus increase anticipatory utility. How voters respond to information is therefore a key determinant of how politicians compete in elections.
 
The analysis reveals that when the consequences of severe climate change are sufficiently large (``catastrophic''), voters maximize anticipatory utility by disregarding signals indicating that the problem is severe. This, in turn, induces politicians to campaign exclusively on policies designed for mild climate change. In contrast, when the welfare losses from severe climate change are moderate rather than catastrophic, multiple equilibria arise. In an efficient equilibrium, voters trust politicians to choose welfare-maximizing policies, thereby creating incentives for politicians to do so. In an inefficient equilibrium, voters expect politicians to ignore relevant information and to campaign on policies appropriate for mild  climate change, which in turn gives politicians incentives to disregard severe climate risks. Thus, when the consequences of severe climate change are moderate, voter trust in welfare-maximizing policy choices becomes self-fulfilling, sustaining politicians' incentives to propose such policies.

Trust in government functions as a coordination device.  When voters believe that politicians will implement policies that match the underlying state, signals are interpreted more faithfully, allowing society to converge to the efficient equilibrium. By contrast, when trust is low---interpreted as the belief that policy choices are largely independent of the underlying state---this link breaks down. In that case, pessimistic voters interpret information in ways that rationalize the policies they expect the government to pursue, giving rise to a self-confirming equilibrium in which inefficient policies persist. The model therefore identifies trust as an important determinant of whether political institutions can translate information about climate risks into effective action.

The analysis shows that an efficient equilibrium is less likely to exist when the anticipated welfare losses from severe climate change are sufficiently large. How the severity of climate change is framed---whether it is presented as a ``catastrophe'' or as a more moderate ``challenge''---may influence voters' expectations about the magnitude of potential losses from climate change. An indirect implication of the model is, therefore, that climate change rhetoric can influence voters' anticipatory utility and, through this channel, their belief formation, which in turn determines equilibrium outcomes. In this sense, highly alarming descriptions need not translate into greater support for policy action and may instead induce defensive belief formation, a pattern emphasized by \citet{ClaytonManningKrygsmanSpeiser:2017} in the quote at the beginning of the paper.

\paragraph{Related Literature:}

The paper contributes to several areas of literature. Firstly, it adds to the literature that examines the incentives of political candidates to select efficient policy platforms when the true state of the world is unknown. Important contributions to this literature include \cite{HeidhuesLagerlof:2003}, \cite{LaslierVanderStraeten:2004}, and \cite{kartik2024information}.
Unlike \cite{HeidhuesLagerlof:2003} and \cite{kartik2024information}, we assume that voters also receive information, which creates the main tension of whether to ignore it or not. In \cite{LaslierVanderStraeten:2004}, voters receive a single public signal as well, whereas in the current paper, each voter receives an independent signal. The two most closely related papers in this literature are \cite{GRATTON:2014} and \cite{CrutzenSisakSwank:2024}. \cite{GRATTON:2014} assumes that candidates observe the state perfectly while voters receive imperfect but informative signals about the state, and the focus of the analysis is to identify conditions that lead to equilibria such that candidates choose  efficient policies. \cite{CrutzenSisakSwank:2024} study a game like \cite{GRATTON:2014}, but there are two groups of voters, elites and commoners, who stochastically differ in their policy preferences and who always differ in  quality of the information they receive. The analysis focusses on deriving conditions such that candidates cater to the elites or engage in populism. Unlike in those papers, the focus of the current paper is on voters' misinterpretation of information and how this influences candidates' incentives to propose efficient policies. This is related to \cite{MillnerOllivierSimon:2020}, who study how confirmation bias affects electoral outcomes. In contrast, we consider a model in which the future affects voters' well-being through anticipatory utility, giving them an incentive to hold motivated beliefs.

As just foreshadowed, a cornerstone of the model is that voters derive anticipatory utility, ``\emph{meaning that the individual experiences pleasant or aversive emotions from thinking about future welfare}'' (\citealp{BenabouTirole:2016}). As in \cite{AkerlofDickens:1982}, \cite{CaplinLeahy:2001}, \cite{BenabouTirole:2002}, \cite{brunnermeier2005optimal}, \cite{Spiegler:2016}, \cite{SpieglerEliaz:2020},  \cite{LittleEtAl:2022}, and \cite{engelmann2024anticipatory}, this means that voters have preferences not only about states and policies but also about their own beliefs. A consequence of this is that voters may engage in motivated reasoning; see, for example, \cite{Kunda:1990}, \cite{BenabouTirole:2016}, \cite{Zimmermann:2020}, \cite{OpreaYuksel:2021}, or \cite{Thaler:2024}. 
The most closely related papers in this area are \cite{LEVY:2014} and \cite{LEYAOUANQ:2023}, which also examine electoral contexts. \cite{LEVY:2014} presents a model featuring a policymaker trying to signal quality/congruency to voters. Voters have imperfect memory and can suppress certain news by purposefully conflating positive and negative signals. In contrast, there is no signaling in the current paper. Furthermore, voters have perfect memory but may choose to interpret some signals as more or less informative than they really are. 
Finally, it is not just voters' beliefs about the state that matter, but also their expectations regarding policy outcomes. This difference introduces a novel self-fulfilling aspect that becomes critical for equilibrium. \cite{LEYAOUANQ:2023} suggests that voters may interpret signals in a contrary manner at a cost. In the current paper, we assume that voters can modulate the perceived informativeness of a signal, but they cannot alter its direction. Moreover, unlike in \cite{LEYAOUANQ:2023}, a key component of our  model is that the state influences baseline welfare levels, which has important consequences for equilibrium.

The paper also contributes to the literature on environmental policymaking. 
\cite{DelfgaauwSwank:2024} show that there can be locked-in effects that prevent the adoption of environmentally friendly policies. \cite{Blumenthal:2024} shows that efficient environmental policies may not be adopted if voters' preferences are expected to change over time. \cite{vanDerStratenEtAl:2024} study the relationship between climate change adoption policies and inequality in society. \cite{BesleyPersson:2023} examine the conditions for green transitions to occur, depending on voters' ideological preferences and the extent to which the production sector remains reliant on brown energy sources. \cite{GULLBERG:2008}, \cite{SHAPIRO:2016}, and \cite{BallesMatterSutzer:2024} show that special interest groups may prevent climate change policies from being adopted. 
In contrast to these papers, the current paper provides an explanation based purely on the electoral incentives of office-motivated politicians in the presence of a common information processing bias among voters.

Finally, the paper contributes to the literature studying trust as a determinant of policy making and welfare. \citet{dasgupta2000trust} argues that higher levels of trust in society reduce transaction costs and thereby increase welfare. Relatedly, \citet{zak2001trust} show that higher trust stimulates investment and economic growth. \citet{aghion2010regulation} demonstrate that lower trust in society causes voters to demand more regulation, and that excessive regulation, in turn, hinders the formation of trust. This mechanism is similar to the current model, where low trust in government leads voters to demand policies not targeted at climate change. However, their model predicts a negative relationship between trust and regulation, whereas in the present paper the relationship between trust and climate policy is positive. This is in line with \cite{Bose:2023}, who shows that higher trust induces politicians to provide more public goods,  in particular, more climate change policies. \citet{besley2024trust} show that state effectiveness is greater when citizens trust their government, because higher trust increases voluntary compliance and thereby reduces implementation costs. \citet{ehrmann2025trust} reviews the literature on trust in central banks and conclude that a high level of trust is an important prerequisite for a successful monetary policy.

\bigskip

The paper is organized as follows. In Section \ref{sec:model}, we describe the baseline model, which is then solved in Section \ref{sec:EQ}.  
In Section \ref{sec:discussion_evidence}, we discuss the main implication of the model. Section \ref{sec:otherextensions} extends the baseline model along several dimensions.  Section \ref{sec:conclusion} concludes. All proofs are relegated to the appendix.

\section{The Model}

\label{sec:model}
In this section we introduce the main building blocks of the model.
There is an unknown state of the world, $\omega\in\{0,1\}$, indicating whether climate change is mild, $\omega=0$, or severe, $\omega=1$. The prior probability that $\omega=1$ is $q\in(0,1)$.
There are two different kinds of actors. On the one hand, there are two  purely office motivated politicians, indexed by $i=1,2$. On the other hand, there is a continuum of voters of mass 1.
Candidates vie for voters' support by proposing policies, $p_i\in\{0,1\}$, to which they commit. Denote the vector of policies by $\mathbf{p}\equiv(p_1,p_2)$.

Voters policy preferences depend on the state of the world $\omega$.
 Given a winning policy $p$ and given a state realization $\omega$, voters' realized policy utility is
\begin{equation}
u(p,\omega)=\left\{\begin{array}{cl}
-p&\text{if }\omega=0,\\
-\Delta-b\left|p-1\right|&\text{if }\omega=1.
\end{array}
\right.
\label{eq:util}
\end{equation}
Hence, for each state realization $\omega$, the optimal policy is $p=\omega$. But the state $\omega$ determines not only optimal policy, but the baseline welfare level in society. In particular,  if $\omega=1$, i.e., when climate change is severe, then baseline welfare decreases by $\Delta>0$, independent of the chosen policy, while it remains constant if climate change is mild, $\omega=0$. The parameter $b>0$ measures the importance of choosing the right policy when climate change is severe: the greater is $b$, the more important is taking the right action.
 We make the following assumption to capture the idea that at the outset, voters believe the optimal policy choice is $p=0$:\footnote{This assumption is not without loss of generality for the model's conclusions. Allowing for $q\geq \frac{1}{1+b}$ would imply that the set of possible equilibria is larger.  We chose to assume $q<\frac{1}{1+b}$, because it seems realistic that voters need to be convinced to support anti-climate change policies. Moreover, as a second benefit, Assumption \ref{assume:q}  keeps the paper's exposition  more concise.}
\begin{assume}
\label{assume:q}
Throughout we  assume that $q< \frac{1}{1+b}.$
\end{assume}

At the beginning of the game, both types of players receive a signal that is informative about $\omega$. In particular, as in \cite{GRATTON:2014}, before announcing policy platforms every politician  $i$ receives a  signal $s_i^P$ that perfectly reveals $\omega$.
Hence, after receiving $s_i^P$, each politician perfectly knows whether climate change is severe or not.

Voters receive two kinds of signals. On the one hand, the policy platforms $\mathbf{p}$ may function as  signals about $\omega$, depending on the strategies chosen by the candidates. On the other hand, they receive information from news or from direct experience about $\omega$. We model this by assuming that each voter receives a signal  $s\in\mathbb{R}$.  When the state is $\omega$, the c.d.f. of $s$ is $\Phi\left(\frac{s-(2\omega-1)\mu}{\sigma}\right)$, where $\Phi$ is the c.d.f. of the standard Gaussian distribution and $\mu>0$ and $\sigma>0$ are parameters. Therefore, the typical monotone likelihood ratio property (MLRP) is satisfied, meaning that any $s>0$ is evidence for $\omega=1$ and increases the belief compared to the prior, while any $s<0$ is evidence for $\omega=0$ and decreases the belief. Following \cite{Callander:2011} and others, we interpret $\mu$ as a measure of the complexity of the issue climate change.
If $\mu$ is small, the issue is very complex, and thus voters tend to hold imprecise beliefs about $\omega$. To the contrary, if $\mu$ is large, climate change is not a very complex issue and beliefs tend to be precise.

 We assume that the complexity of the issue climate change is such that, absent motivated beliefs and any signaling about the state $\omega$ through platform choices $\mathbf{p}$, the election aggregates information whenever $p_1\neq p_2$. In other words: if both policies are offered, a majority chooses to vote for the welfare-maximizing policy. The following assumption guarantees that this is true:

\begin{assume}
\label{assume:psi_BAYES}
 Throughout we  assume that \[
\mu\geq \hat{\mu}\equiv\frac{\sigma  \sqrt{\ln \left(\frac{1-q}{b  q}\right)}}{\sqrt{2}}.
\]
\end{assume}

 We assume that voters derive anticipatory utility. In particular, a voter has expectations about her future utility, which is derived from a belief bout the true state and a second belief about the enacted policy. This causes anticipatory utility, ``\emph{meaning that the individual experiences pleasant or aversive emotions from thinking about future  welfare}'' \citep{BenabouTirole:2016}. To increases anticipatory utility (to be defined precisely below), a voters may use motivated reasoning, which means  that she may update beliefs using a \emph{distorted} complexity parameter $\hm$. Clearly, processing information using such  a distorted complexity parameter comes at a cost, because higher anticipatory utility may imply lower utility because of imperfect decision making. As \cite{BenabouTirole:2016} write, ``\emph{one can react to bad news objectively, which leads to better decisions but
having to live with grim prospects for some time [\dots], or adopt a more ``defensive'' cognitive response that makes life easier until the day of reckoning,
when mistakes will have to be paid for.}''  We do not model these costs in the baseline model to be able to identify the pure effect of motivated reasoning, but in Section \ref{sec:costly} two models of distortions are discussed. If a voter is indifferent between having $\hm=\mu$ and some other $\hm\neq \mu$, we assume without loss of generality that she chooses $\hm^*=\mu$.
Choosing $\tilde{\mu}>\mu$ implies the voter interprets the signal as more informative than it really is, and hence beliefs will change excessively,  while choosing $\tilde{\mu}<\mu$ implies a more conservative stance and that beliefs move less than they should.

Based on $s$ and $\mathbf{p}$, and after choosing $\tilde{\mu}$, each voter forms beliefs $\pi(s,\mathbf{p},\hm)$ and $\kappa$. $\pi(s,\mathbf{p},\hm)$ is the posterior belief about $\omega$. $\kappa$ denotes the perceived probability that policy~1 is implemented, given voters' strategies and beliefs and the induced distribution of signals. Because candidates commit to policy platforms, clearly $\kappa=1$ if $p_1=p_2=1$, and $\kappa =0$ if $p_1=p_2=0$, and thus $\kappa$ is uniquely determined by $\mathbf{p}$ if $p_1=p_2$. If $p_1\neq p_2$, $\kappa$ can be related to $\pi$, but it can also be independent, for example, if a voter holds the expectation that independent of the state $\omega$, the policy that will be implemented is $p\in\{0,1\}$. 

Equipped with these beliefs, we can now  calculate a voter's anticipatory utility:
\begin{equation}
\label{EQ:AU}
AU(s,\mathbf{p},\hm)=-\kappa \left[\pi(s,\mathbf{p},\hm) \cdot \Delta+ (1-\pi(s,\mathbf{p},\hm)) \right]-(1-\kappa)\left[\pi(s,\mathbf{p},\hm)  \cdot(\Delta+b)\right]
\end{equation}
Thus, anticipatory utility equals the utility the voter expects to receive once policies are determined.

Note that a voter is never pivotal, because there is a continuum of voters.  We thus assume that voters
vote sincerely.
%as defined by \cite{AustenSmithBanks:1996}. 
In  words, each voter forms beliefs based on $s$,  $\mathbf{p}$, and $\hm$ and then votes to maximize \eqref{eq:util}.

 We are now in a position to define equilibrium of our game.
(i) each voter maximizes $AU(s,\mathbf{p},\hm)$ given  a belief $\kappa$ about the probability that policy 1 will be enacted;  (ii) each voter votes sincerely given $s$ and $\hm$; (iii) $\kappa$ is correct  given $s$, $\hm$, and given the strategies of all other voters; (iv) politicians choose policy platforms that maximize the probability to be elected, given their expectations about voters' behavior. As typical in signaling games, there are multiple equilibria.  We focus attention on \emph{symmetric equilibria}, that is, equilibria in which candidates, holding the same information and trying to achieve the same, choose identical strategies. Further, if on the equilibrium path both candidates choose the same pure strategy, then we must have $p_1=p_2$. If, off the equilibrium path, voters observe $p_1\neq p_2$, we assume that they do not learn anything from this about $\omega$. That is, if off the equilibrium path $p_1\neq p_2$, then observing $\mathbf{p}$ does not change voters'  belief  about $\omega$.\footnote{This could be justified, for example, because candidates have types as well. An ideological candidate always chooses $p_i=i-1$, $i\in\{1,2\}$. A strategic candidate chooses the policy that is expected to maximize the probability to win the election. A model in which each candidate $i$ is ideological with probability $\gamma\in(0,1)$ and non-ideological with probability $1-\gamma$ would yield the same results regarding the platform choices of office-motivated candidates as the assumption of non-informative deviations.}

\section{Equilibrium}
\label{sec:EQ}
In this section, we solve the baseline game.  We begin by analyzing equilibrium play in the voting subgame. In Section \ref{ssec:candidates}, we analyze candidates' incentives and equilibrium behavior, and in Section \ref{ssec:equilibrium} we study the equilibrium of the whole game.

\subsection{Voter Behavior}
First we turn to the optimal motivated beliefs of voters.
The focus is on situations in which  $p_1\neq p_2$, because otherwise voters choose a winner, but not a winning policy. If $p_1\neq p_2$ off the equilibrium path,  $\mathbf{p}$ is not informative about $\omega$.
However, it may be so on the equilibrium path. In a symmetric equilibrium,  $p_1\neq p_2$ is only possible in mixed strategies. Because both candidates hold the same information, assume that, when the state is $\omega$, each chooses $p=1$ when  with probability $\rho_\omega$ and $p=0$ with probability $1-\rho_\omega$. Therefore, if $\rho_1=1$ and $\rho_0=0$, then both choose the policy that matches the state with probability 1. If $\rho_0=\rho_1=0$, then candidates never choose $p=1$, irrespective of $\omega$.

If  $p_1\neq p_2$ off-equilibrium, $\mathbf{p}$ is not informative by assumption, and  $p_1\neq p_2$ in a symmetric pure strategy equilibrium is not possible. 
Hence, the belief about $\omega$ of a voter who receives signal $s$ and who chooses $\hm$, is

\begin{equation}
\label{eq:pi}
\pi(s,\mathbf{p},\hm)=\dfrac{q \rho_1(1-\rho_1)}{q\rho_1(1-\rho_1)+(1-q)\rho_0(1-\rho_0)e^{-\frac{2 \hm  s}{\sigma^2}}}.
\end{equation}
Note that this belief is not properly defined when one of the mixed strategies becomes a pure strategy, i.e., when $\rho_\omega\in\{0,1\}$.\footnote{The reason is that there cannot be a pure symmetric pure strategy equilibrium in which candidates choose $p_1\neq p_2$.} In a symmetric pure strategy equilibrium, platforms are perfectly informative about $\omega$.

To see how a voter optimally chooses her motivated belief $\hm$, take the derivative of  $AU(s,\mathbf{p},\hm)$ with respect to $\hm$:
\begin{equation}
\label{eq:dAUdsigma}
\dfrac{\partial  AU(s,\mathbf{p},\hm)}{\partial \hm}=-
\frac{2 (1-q) q (1-\rho_0) \rho_0 (1-\rho_1)
   \rho_1 s \left[b +\Delta-(b +1) \kappa  \right] e^{\frac{2 \mu 
   s}{\sigma ^2}}}{\sigma^2\left((1-q) (1-\rho_0) {\rho_0}  +q
   (1-\rho_1) \rho_1  e^{\frac{2 \mu  s}{\sigma
   ^2}}\right)^2}
\end{equation}
If $\kappa=\tilde{\kappa}\equiv\frac{b+\Delta}{b+1}$, then this is equal to zero, independent of $s$. If $\Delta\geq 1$, then $\tilde{\kappa}\geq1$.
%and because $\kappa$ is a belief, the only possibility is $\kappa\leq\tilde{\kappa}$, implying that the sign of \eqref{eq:dAUdsigma} is independent of  $\kappa$. 
In particular, if this is the case, then  $AU\kps$ weakly decreases in $\hm$ if $s>0$, and it weakly increases in $\hm$ if $s<0$.
Otherwise, that is, if $\Delta< 1$, then whether or not $AU\kps$ increases in $\hm$ depends on both $s$ and the belief $\kappa$. 

Recall that $\kappa$ captures the belief  about the policies that will be enacted. It may depend on a voters signal if the voter beliefs the state $\omega$ to be relevant for the winning chances of the two candidates. If a voter believes that candidate that offers $p=\omega$ gets elected, her belief will equal  $\kappa=\pi(s,\mathbf{p},\hm)$. For example, if all voters vote informatively this would be the case.
Such a situation can be described as the voter trusting the government to take the correct action. 
However, this needs not be the case, and a voter may hold the expectation that either $p=0$ or $p=1$ are chosen independent of the realization of $\omega$.  
For example, if there is little trust in government regarding policy choices, then $\kappa\neq\pi(s,\mathbf{p},\hm)$, and if trust is very low, we have $\kappa\in\{0,1\}$.

Lemma \ref{lem:sigma} states for each voter the optimal  distortion $\hm^*$ as a function of $s$ and possibly $\kappa$:

\begin{lemma}
\label{lem:sigma}
Let $\tilde{\kappa}\equiv\frac{b+\Delta}{b+1}$.
\begin{enumerate}
\item[(a)] If $s=0$ or $\kappa=\tilde{\kappa}$, then  $\hm^*=\mu$ for all $\Delta$.
  \item[(b)] If $\Delta>1$, then (i) $\hm^*=\infty$ if $s<0$, and (ii) $\hm^*=0$, if $s>0$.
  \item[(c)] If $\Delta=1$ and  $\kappa<1$,  (i) $\hm^*=\infty$, if $s<0$, and (ii) $\hm^*=0$, if $s>0$.
  \item[(d)] If $\Delta<1$, then  (i) $\hm^*=\infty$, if either [$s<0$ and $\kappa<\tilde{\kappa}$] or [$s>0$ and $\kappa>\tilde{\kappa}$],  and (ii) $\hm^*=0$, if either [$s<0$ and $\kappa>\tilde{\kappa}$]  or [$s>0$ and $\kappa<\tilde{\kappa}$].
\end{enumerate}
\end{lemma}

The lemma provides us with an important intermediate result.  We see that beliefs about policy only matter if severe climate change causes only moderate baseline welfare losses. Otherwise,  if severe climate change leads to catastrophical welfare losses, then \textit{any} signal indicating that $\omega=1$ will be interpreted as pure noise. To the contrary, \textit{any} signal indicating $\omega=0$ will be accepted as a perfect indication that climate change is indeed mild, independent of the signal's real strength.  
%Finally, a voter with a signal $s=0$ or a voter with $\kappa=\tilde{\kappa}$ has no incentive to distort her signal's precision and hence updates her belief about $\omega$ like a perfect Bayesian.

 What does this imply for the beliefs about $\omega$ the voters hold? Clearly, if $\hm=0$, then 
\[\pi(s,\mathbf{p},0)= \frac{q\rho_1(1-\rho_1)}{q\rho_1(1-\rho_1)+(1-q)\rho_0(1-\rho_0)}.
\]
This is the same belief a Bayesian voter observing $s=0$ holds. Moreover, if  $\hm=\infty$, then any signal that contains only the slightest bit of information will completely move beliefs to the extremes, and hence  $\pi(s,\mathbf{p},\infty)\in\{0,1\}$. Only if $\kappa=\tilde{\kappa}$ will the belief be a non-constant continuous function of $s$ and it equals $\pi(s,\mathbf{p},\mu)$:

\begin{coroll}
Let $\hat{\pi}\equiv \frac{q\rho_1(1-\rho_1)}{q\rho_1(1-\rho_1)+(1-q)\rho_0(1-\rho_0)}$.
\begin{enumerate}
  \item[(a)] If $\Delta>1$, then
  \[
  \pi(s,\mathbf{p},\hm^*)=\left\{
  \begin{array}{cl}
  \hat{\pi}&\text{if }s\geq 0\\
  0&\text{if }s< 0
  \end{array}
  \right.
  \]
  \item[(b)] If $\Delta= 1$,
  then
  \[
  \pi(s,\mathbf{p},\hm^*)=\left\{
  \begin{array}{cl}
    \pi(s,\mathbf{p},\mu)&\text{if }\kappa=1\\
  \hat{\pi}&\text{if }s\geq 0\wedge \kappa<1\\
  0&\text{if }s< 0 \wedge \kappa<1
  \end{array}
  \right.
  \]
  \item[(c)] If $\Delta< 1$,
  
  \[
  \pi(s,\mathbf{p},\hm^*)=\left\{
  \begin{array}{cl}
   \pi(s,\mathbf{p},\mu)&\text{if }\kappa =\tilde{\kappa}\lor s=0\\
  0&\text{if }s< 0\wedge \kappa <\tilde{\kappa}\\
  \hat{\pi}&\text{if }\left(s> 0\wedge \kappa <\tilde{\kappa}\right)\vee \left(s< 0\wedge \kappa >\tilde{\kappa}\right)\\
  1&\text{if }s> 0\wedge \kappa >\tilde{\kappa}
  \end{array}
  \right.
  \]
\end{enumerate}
\label{cor:beliefs}
\end{coroll}

 We now know the beliefs of all voters as functions of their signals and of their policy belief $\kappa$. Corollary \ref{cor:beliefs} shows that if $\Delta> 1$, and hence severe climate change has catastrophic consequences, then the equilibrium has a simple structure, and $\kappa$ actually plays no roll. Voters have two different beliefs, $\pi=0$ and $\pi=\hat{\pi}$.  What does this imply for voters' decisions at the ballot? Recall that voters vote sincerely for the alternative that they believe maximizes \eqref{eq:util}. The expected utility from policy 1 is $u(p=1)=-\pi \Delta-(1-\pi)$, while from policy 0 she gets $u(p=0)=-\pi (\Delta+b)$.
Hence, the voter cast her ballot for policy 1 iff
\begin{equation}
\label{eq:voting_rule}
u(p=1)>u(p=0)\Leftrightarrow \pi\kps >\tilde{\pi}\equiv\frac{1}{1+b}.
\end{equation}
If the reverse is true, then policy 0 is strictly preferred, and if $\pi=\tilde{\pi}$, then a voter is indifferent.
Clearly, $1>\tilde{\pi}>0$. Note that by Assumption \ref{assume:q} we have $\tilde{\pi}>q$.

If $\Delta> 1$,  then voters hold two different beliefs, either $\pi=0$ or $\pi=\hat{\pi}$. A voter with the former belief always votes for policy 0, while the decision at the ballot of the other voter depends on $q$, $b$, as well as on the politicians' strategies $\rho_0$ and $\rho_1$. If $\hat{\pi}>\tp$, then these voters vote for policy 1, and as a consequence all voters vote informatively, i.e., they vote according to their signals. This means that in each state $\omega$ the policy that matches this state is chosen. If, however, $\hat{\pi}\leq\tp$, then a majority of voters always supports policy 0, implying it wins independent of the state. If $\Delta=1$, beliefs are the same except when $\kappa=1$, in which case a voter holds a  Bayesian belief about $\omega$ and votes accordingly.

Things are slightly different when $\Delta<1$, as now $\kappa$ starts to matter. As before, a situation in which all voters hold a belief $\kappa<\tilde{\kappa}$, leading to the above discussed situation of all voters voting for policy 0 if faced with a choice, continues to exist.  On the other hand, if  $\hat{\pi}\geq\tp$ and $\kappa>\tilde{\kappa}$ for sufficiently many  $s$, then policy 1 wins independent of the state.

 What about an efficient outcome, in which voters collectively choose the policy that matches the state $\omega$?  Assume each voters wants to believe her signal and thus has $\kappa<\tilde{\kappa }$ when $s<0$ and $\kappa>\tilde{\kappa }$ when $s>0$. Then every voter votes informatively, and therefore the policy matching the state  wins for all $\hat{\pi}\in(0,1)$.
Hence, if even severe climate change decreases baseline welfare only moderately, then there  may exist an equilibrium in which voters vote informatively, and hence the correct policy is chosen with probability 1.

 We summarize the results of this section in our next proposition:
\begin{prop}
\label{prop:voting}
Assume $p_1\neq p_2$.\begin{enumerate}
  \item If $\Delta\geq 1$, then there is a unique  equilibrium.
  \begin{enumerate}
    \item If $\hat{\pi}\leq \tp$, then $\kappa^*=0$ for all $s$ and a majority of voters  votes for policy 0.
    \item If $\hat{\pi}> \tp$, then $\kappa^* =0$ if $s< 0$ and $\kappa^* =\hat{\pi}$ if $s\geq 0$, and a majority  votes for  $p=\omega$.
  \end{enumerate}
  \item If $\Delta< 1$, then there  exist multiple  equilibria.
  \begin{enumerate}
    \item For any $\hat{\pi}\in(0,1)$, there exists a subgame equilibrium  with  $\kappa^* =0$ if $s< 0$ and $\kappa^* =1$ if $s> 0$, and a majority of voters always votes for  policy $p=\omega$.
    \item For $\hat{\pi}\geq\tilde{\pi}$, there exists a subgame equilibrium with $\kappa^*=1$ and policy 1 always wins.
    \item For $\hat{\pi}\leq\tilde{\pi}$, there exists a subgame equilibrium with $\kappa^*=0$ and policy 0 always wins.
    \item There is no subgame equilibrium in which   policy $p\neq \omega$ always  wins.
  \end{enumerate}
\end{enumerate}
\end{prop}

There always exists a subgame equilibrium in which information aggregates, if severe climate change decreases baseline welfare not too much, i.e., if $\Delta$ is small. However, this equilibrium depends on the expectations about the efficiency of the political process. If voter believe that the policy suggested by their signal is chosen, then the election leads to efficient results. But if voters are pessimistic, and believe policy 0 will be chosen no matter what is the true state $\omega$, then this becomes a self-fulfilling prophecy if $\hat{\pi}$ is low.  When $\Delta$ becomes larger, and hence severe climate change leads to more dire welfare losses, then the efficient equilibrium disappears when $\hat{\pi}$ is small.

\subsection{Candidates}
\label{ssec:candidates}

 We now study the platform choices of the candidates as a function of their signals.
Of course, when they decide which policies to offer, they reason forward thinking about which policy is more likely to lead to electoral success. Therefore voters' reactions to policy choices, in particular the one's formalized in Proposition \ref{prop:voting}, matter for candidates' incentives.

From the perspective of an office motivated candidate, the optimal platform is the one that maximizes the chance to win the election. In a symmetric equilibrium, if both candidates choose identical platforms, voters may learn from the platforms' congruence something about the true state $\omega$, but they cannot choose policies anymore, since all candidates offer the same. Because voters only care about policies,  both candidates win with an equal probability of 50 percent if $p_1=p_2$.
 If some candidate deviates and chooses off-equilibrium a platform different from his opponent, voters learn nothing about $\omega$ from this deviation. However, if equilibrium play permits  $p_1\neq p_2$, then
  platforms may indeed be partially informative and change voters' beliefs, as discussed before.

First consider potential mixed-strategy equilibria. Focussing on symmetric equilibria, the probability that each candidate choose
policy 1 in state $\omega$ is $\rho_\omega$ as before. If $\rho_1\neq \rho_2$,  voters learn from platform choices even if $p_1\neq p_2$. Choosing such a mixed strategy can only be an equilibrium if it leads to a chance of winning of 50\%, because candidates need to be indifferent. The probability to win is 50\%, if either both receive exactly half the votes, or if the probability that a majority chooses either candidate is 50\%. But note that candidates know the state $\omega$, and thus they can infer the exact distribution of signals voters receive. Therefore, in a mixed-strategy equilibrium we need to have that both receive exactly half the votes.
This implies policy 1 is implemented with probability of 50\%, independent of the state $\omega$. Hence, we need to have $\kappa=\ot$ for all $s$. If $\tilde{\kappa}<\ot$, any voter with $s>0$ will hold belief $\pi=1$ and thus they all vote for  1. But this means that in state $\omega=1$, the candidate offering $p=1$ wins with certainty, and thus this cannot be equilibrium. If $\tilde{\kappa}>\ot$, any voter with $s<0$ will hold belief $\pi=0$ and thus they all vote for policy 0. But this means that in state $\omega=0$ the candidate offering $p=0$ wins with certainty, and thus this cannot be equilibrium, either. Finally, if $\tilde{\kappa}=\ot$, any voter will choose $\hm^*(s)=\mu$. But then information aggregates by Assumption \ref{assume:psi_BAYES}, and thus  also this cannot be  an equilibrium belief. Hence, we can conclude that no mixed strategy equilibrium can exist.

 What about pure strategy equilibria? In a symmetric equilibrium, we must have $p_1=p_2$, because candidates hold the same information. Suppose  voters observe off-equilibrium that $p_1\neq p_2$ and  both  candidates expect voters to vote informatively. Hence,  they vote for policy 1 if $s>0$, for policy 0 if $s<0$, and they randomize when $s=0$. Then, in each state $\omega$, the candidate offering the policy that matches the state wins the election. Hence, when choosing which policy platform to offer,  candidate $i$ knows that $p_i=\omega$ wins the election if $p_{-i}\neq \omega$ and otherwise both win with a probability of 50\%. To the contrary, choosing $p_i\neq\omega$ wins the election with 50\% if also $p_{-i}\neq \omega$ and otherwise loses the election for sure. Hence,   $p_i=\omega$ is the unique optimal strategy. It follows that if sufficiently many voters vote informative, candidates will choose informative policy platforms. To the contrary, if candidates expect a sufficiently large number of voters to not vote informatively when $p_1\neq p_2$, then there exists  a policy $\hat{p}\in\{0,1\}$ that wins the election with certainty, independent of the state $\omega$.
Candidates then know they win the election by choosing $p_i=\hat{p}$ if $p_{-i}\neq \hat{p}$, and they win  with a probability of 50\% if $p_{-i}=\hat{p}$. Choosing $p_i\neq \hat{p}$ will lose the election for sure if $p_{-i}=\hat{p}$, and hence both candidates are best served by offering $\hat{p}$.

This leads to the next  result:

\begin{prop}
\label{prop:cands}For given $\hm^*(s)$ for all $s$, there is a unique symmetric equilibrium in the platform choice stage, which is in pure strategies. Candidates choose $p_i^*=\omega$, if they expect a majority of voters to vote for the policy that matches the state, when $p_1\neq p_2$.
Otherwise,  there exists a policy $\hat{p}$ that is expected to win independent of the realization of $\omega$ if $p_1\neq p_2$, and  both candidates choose $p_i^*=\hat{p}$.
\end{prop}

\subsection{Equilibrium Policy Platforms}
\label{ssec:equilibrium}
 We can now
%that we have established candidates' equilibrium choices a s function of expected voter choices at the ballot,
determine equilibrium platform choices as functions of the parameters of the game. If severe climate change leads to catastrophic baseline welfare level losses,  $\Delta\geq1$, then voters ignore any $s>0$. Our analysis reveals that then a majority always votes for policy 0, whenever $p_1\neq p_2$. Proposition  \ref{prop:cands} tells us that in this case candidates never campaign on policy $p=1$, independent of the information they hold. Hence, if severe climate change is catastrophic, then candidates will ignore any information and always choose the optimal policy for mild climate change, $\mathbf{p}^*=(0,0)$.

 When $\Delta< 1$, the set of equilibria becomes larger. The reason is that now voters' beliefs about the enacted policy become self-fulfilling. If voters trust the government  in the sense that they believe the policy that is optimal given $\omega$ wins, then voters always vote informatively,  implying that  $\mathbf{p}^*=(\omega,\omega)$ is indeed an equilibrium. However, there always coexists the equilibrium in which  which candidates  ignore their information  and  choose $\mathbf{p}^*=(0,0)$.

The next proposition states the paper's main result and formalizes the above intuitions:
\begin{prop}
If $\Delta\geq1$, then there exists a unique \ps equilibrium in which  $\kappa^*=0$ for all $s$ and  candidates  choose $\mathbf{p}^*=(0,0)$.
If instead $\Delta<1$, then there exist multiple equilibria. In particular, there exists one pure strategy equilibrium in which, for each $s$,  $\kappa^*=\pi(s,\mathbf{p},\infty)$ and candidates choose  $\mathbf{p}^*=(\omega,\omega)$, and there exists another  pure strategy  equilibrium in which  $\kappa^*=0$ for all $s$ and candidates choose  $\mathbf{p}^*=(0,0)$.
\label{prop:inactivity}
\end{prop}
The analysis shows that when motivated beliefs are present, a self-fulfilling efficient equilibrium exists as long as the worst consequences of climate change are not too severe, i.e., $\Delta < 1$. In this equilibrium, each voter takes her signal at face value and therefore votes informatively. As a result, candidates have the correct incentives and campaign on the policy that is optimal given $\omega$. Consequently, the probability that the winning policy matches $\omega$ is 1.

However, there also always exists a \textit{bury-your-head-in-the-sand} equilibrium, independent of $\Delta$. In this equilibrium, voters choose to ignore any signal $s > 0$, while treating any signal $s < 0$ as conclusive evidence that $\omega = 0$. The ex ante probability that the correct policy is chosen in this equilibrium is $1-q$.

%XXX 
%"Proposition 3. If Δ ≥ 1, then there exists a unique pure strategy equilibrium in which κ* = 0 for all s and candidates choose p* = (0,0). If instead Δ<1, then there exist multiple equilibria. In particular, there exists one pure strategy equilibrium in which, for each s, κ* = π(s,p,∞) and candidates choose p* = (ω,ω), and there exists another pure strategy equilibrium in which κ* = 0 for all s and candidates choose p* = (0,0)."

%\verbatim{Message: At first the expression κ* = π(s,p,∞) in the efficient equilibrium (Δ<1) made me think that voters' on-path belief about the winning policy might still depend on their private signal.  After tracing back the definition of π(·) and Proposition 2, we realised that once candidates play p* = (ω,ω) the platform vector itself fully reveals the state, so π(s,p,∞) collapses to ω and κ* is actually constant.  You might consider stating this simplification explicitly—for example by adding “(so π(s,p,∞)=ω and therefore κ*=ω on the equilibrium path)”—to forestall momentary confusion about whether κ* varies with s.}

\section{Discussion and Implications}

\label{sec:discussion_evidence}

In this section, we discuss two important implications of the model: that the efficiency of policy-making in the face of climate change depends on both  \emph{rhetoric} and \emph{trust in government}.
%Suppose that, without additional information, voters are leaning towards a policy that is adequate for mild climate change, $q<\tilde{q}$.

\subsection{Trust in Government}
\label{ssec:TIG}

 When anticipated welfare losses from severe climate change are sufficiently large, the model admits a unique, inefficient equilibrium. By contrast, when these losses are moderate, multiple equilibria arise. In this case, equilibrium selection depends critically on voters' expectations about how policy choices relate to the underlying state of the world, which we interpret as trust in government.

Following \citet{cabral2005economics} and \citet{besley2024trust}, we define trust as voters' expectations that political institutions implement policies that are responsive to the true state. In the model, voters hold two distinct beliefs: a belief $\pi$ about the state of the world and a belief $\kappa$ about the probability that policy~1 will be implemented. In the efficient equilibrium with moderate welfare losses, voters' beliefs about the state take extreme values. As shown in Section~3, motivated belief formation implies that $\pi \in \{0,1\}$, with $\pi=1$ when the voter's signal satisfies $s>0$ and $\pi=0$ when $s<0$.

In this environment, trusting the political process means that voters expect policy choices to mirror the realized state. Since policy~1 is optimal if and only if climate change is severe, a voter who believes $\pi=1$ expects policy~1 to be implemented with probability one, whereas a voter who believes $\pi=0$ expects policy~1 to be implemented with probability zero. It therefore follows directly in this equilibrium that the perceived probability that policy~1 is implemented coincides with the belief that the state is severe, so that $\kappa=\pi$.

 When welfare losses from severe climate change are moderate and voters have  trust, they optimally take their signals at face value and vote \textit{informatively}. Anticipating this behavior, politicians have incentives to follow their information and propose welfare-maximizing policies, thereby sustaining the efficient equilibrium. Trust thus operates as a coordination device: by aligning voters' expectations about policy implementation with their beliefs about the state, it ensures that information is processed and aggregated efficiently.

Conversely, when trust is low, voters expect policy choices to be weakly related or unrelated to the underlying state, so that $\kappa$ becomes insensitive to information contained in signals. Such pessimistic expectations weaken incentives to process information accurately and can become self-confirming. Anticipating that voters will not respond informatively, politicians optimally ignore their own information and campaign on policies designed for mild climate change, even when climate change is severe. In this way, low trust sustains an inefficient equilibrium.

The model therefore implies a positive relationship between trust in government and the likelihood that electoral competition leads to effective climate policy, provided that welfare losses from severe climate change are not too large. This implication is consistent with Figure \ref{fig:scatter} and recent empirical evidence showing that higher trust in political institutions is associated with greater provision of public goods, including climate policy \citep{Bose:2023}.

Importantly, this relationship arises only through the interaction between trust and motivated belief formation; absent motivated reasoning, voters' beliefs about the state would be independent of expectations about policy implementation, and trust would not affect equilibrium policy choices.

\subsection{The Importance of Rhetoric}
A first implication concerns political rhetoric. In the model, an efficient equilibrium is less likely to exist when the anticipated welfare losses from severe climate change are sufficiently large. The reason is that larger anticipated losses increase anticipatory disutility and thereby strengthen incentives for voters to engage in defensive belief formation, which undermines coordination on the efficient outcome.

An indirect corollary is that the way climate risks are described may matter for equilibrium selection. To the extent that rhetoric affects voters' expectations about the magnitude of potential welfare losses from climate change, it can shift the strength of anticipatory disutility and, through this channel, the extent of belief distortion. In this sense, highly alarming descriptions need not translate into greater support for policy action and may instead trigger defensive responses, echoing the concern raised by \citet{ClaytonManningKrygsmanSpeiser:2017}.

This argument does not require taking a stand on any particular model of communication. One interpretation is that rhetoric operates through framing, in the sense that equivalent information can induce systematically different perceptions depending on how outcomes are described \citep{tversky1981framing,tversky1989rational,ellingsen2012social,spiegler2014competitive}. Another interpretation is that rhetoric is informative, in line with cheap-talk and persuasion frameworks in which language conveys signals about states or consequences \citep{CrawfordSobel:1982,kartik2019informative,KamenicaGentzkow:2011,AlonsoCamara:2016}. The present model highlights a simple channel common to both interpretations: by shaping perceived severity, rhetoric can affect anticipatory utility and thus the mapping from information to policy outcomes.

\section{Extensions}
\label{sec:otherextensions}
In this section we informally discuss  further extensions that seem relevant. For the sake of brevity and clarity,  throughout this section we  assume that $\Delta>1$. In Section \ref{sec:costly} we study the implication of costly deviations from Bayesian information processing. In  Section \ref{ssec:policymotiv} we discuss how conclusions  change if candidates are not only office motivated, but care also about welfare.

\subsection{Costly Distortions}
\label{sec:costly}
In this section, we extend the previous analysis by accounting for the costs that a voter may incur when distorting the issue's complexity. Introducing costs allows us to scrutinize the above predictions about how  beliefs may be distorted, and it enables us to assess the robustness of the earlier results on policy choices.

 We distinguish between two types of costs. First, we analyze utility costs from making bad decisions (\citealp{BenabouTirole:2002,BenabouTirole:2016,brunnermeier2005optimal}).  We then study purely cognitive  costs from choosing distorted beliefs   (\cite{LEVY:2014,LEYAOUANQ:2023,engelmann2024anticipatory}). For proofs of the statements, see Appendix \ref{app:costly}.

\subsubsection{Cost from Imperfect Decision Making}
Assume each voter  places a weight $\lambda \in[0,1]$ on  anticipatory utility and  $1-\lambda$ on the expected utility loss relative to the Bayesian-optimal action. In particular, voters aim at maximizing $ W^{UC}=\lambda AU+(1-\lambda)C^U$, where $C^U\leq 0$ is the utility cost of deviating from Bayesian information processing.

If distortion does not affect choice, it has effectively zero utility cost. Because $q<\frac{1}{1+b}$, a sufficiently positive $s$ is needed to change Bayesian behavior from voting for $0$ to voting for $1$. Denote the minimum signal necessary by $\tilde{s}>0$. Anybody with only a weak signal indicating that climate change is severe, $s\in(0,\tilde{s})$, or a signal indicating that climate change is mild, $s\leq0$, will, even as a Bayesian, vote for policy $0$. Hence, for these voters, choosing $\tilde{\mu} = \infty$ when $s<0$ and $\tilde{\mu} = 0$ when $s\in(0,\tilde{s})$ remains optimal.

If the distortion does change behavior, it has an expected cost. This cost is the difference between expected utility under Bayesian beliefs with the optimal action and the expected utility under Bayesian belief with the suboptimal action. The only voters for whom distortion may change the action are those with sufficiently high signals, $s>\tilde{s}$. 

 When deciding if and how to distort beliefs, a voter faces two potentially optimal choices: (i)  $\tilde{\mu} = 0$, and thus interpreting any signal as irrelevant; and (ii) choosing the smallest possible $\tilde{\mu}$ such that the choice at the ballot is not altered but anticipatory utility increases.\footnote{This option strictly dominates $\tilde{\mu}=\mu$.} 
The voter's welfare from (i) is
\[
 W^{UC}(\tilde{\mu}=0) = - q(b+\Delta)\lambda-\left(b - \frac{(1-q)(1+b)}{1 -\big(1 - e^{\frac{2s\mu}{\sigma^2}}\big)q}\right)(1-\lambda).
\]
To determine the utility from (ii), note that the belief about $\omega$ given this must equal $1/(1+b)$. Thus,
\[
 W^{UC}(\tilde{\mu}=\mu_c)  = -\frac{(b+\Delta)\lambda}{1+b}.
\]

 When comparing these two welfare levels, note that when $s$ is such that a Bayesian voter is indifferent between voting for policy $0$ or policy $1$, the utility cost from distortion is zero, and thus $ W^{UC}(\tilde{\mu}=0) >  W^{UC}(\tilde{\mu}=\mu_c)$. Moreover, $ W^{UC}(\tilde{\mu}=0)$ is strictly decreasing in $s$. The reason is that if a voter has a very strong signal, the Bayesian belief is very strong, and the utility cost from acting against this belief is high. Therefore, if $ W^{UC}(\tilde{\mu}=0) > W^{UC}(\tilde{\mu}=0)$ even when $s \to \infty$, then any voter with $s>0$ will choose $\tilde{\mu}=0$. This is true if $\lambda$ is sufficiently large. In particular, if
\[
\lambda >  \frac{b(1+b)}{b\big(2-q+(1-q)b\big) + \big(1-q-qb\big)\Delta}\in(0,1).
\]
Otherwise, for lower levels of $\lambda$, a voter with a very strong signal will choose the non-consequential distortion $\mu_c$ and vote according to the signal. Any other voter with $s>0$, however, chooses $\tilde{\mu}=0$, thereby ignoring the signal and voting for policy $0$.
Hence, the extended model of motivated beliefs provides a more nuanced prediction of voter behavior. If $\lambda$ is large, introducing costs changes nothing. However, if $\lambda$ is sufficiently small, then a strong signal will be interpreted as evidence of severe climate change, and voters receiving such a signal will vote for policy~1. Denote the threshold for a sufficiently strong signal by $\hat{s}$ and note that $\hat{s}>\tilde{s}$.

To shift the equilibrium toward one in which politicians offer policies for severe climate change, this is not yet sufficient. The reason is that if $\mu$ is small---i.e., if climate change is a highly complex issue---then only few voters receive $s>\tilde{s}$. Thus, in this case, there exists a unique equilibrium in which no politician ever offers policy~1. However, for large enough $\mu$, more than 50\% of voters receive $s>\tilde{s}$ in state 1, and thus a majority votes for policy~1. This creates incentives for politicians to follow their information and offer policies targeting severe climate change.

\subsubsection{Cognitive Costs}
Next consider pure cognitive costs from distorting information processing. Distorted information processing can lead to cognitive costs because, for example, sustaining self-serving or overly optimistic beliefs may conflict with one's deeper desire for accuracy and coherence, creating dissonance and reducing overall  well-being. Hence, in the following we assume that there is a cost of $C^{C}(\hm,\mu)$ associated with choosing $\hm$.

\begin{figure}
  \centering
  \subfigure{
\includegraphics[width=.450\textwidth]{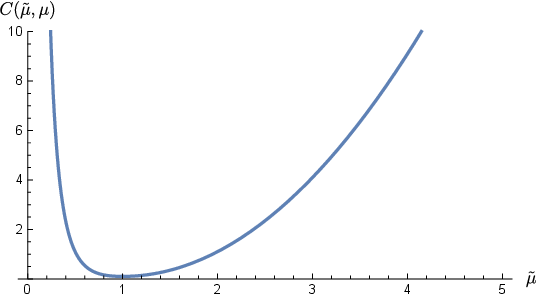}
}
\subfigure{
\includegraphics[width=.450\textwidth]{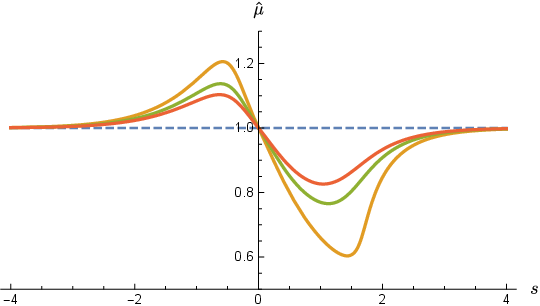}
}
\caption{Left panel:  Cost function when $\mu = 1$ and $c = 2$. Right panel: $\hm$ as a function of $s$ for $\mu=1$ and different values of $c$. Higher $c$ moves $\hm$ closer to $\mu$.}
\label{fig:cost_distortion}
\end{figure}

How should such a cost function look like?
First, it should be convex increasing in $|\mu-\tilde{\mu}|$ and zero when $\tilde{\mu}=\mu$.  Moreover, it should \emph{not} be symmetric in the sense that $\hm=\mu+\epsilon$ and $\hm=\mu-\epsilon$ should not lead to the same cost.\footnote{To see why, assume $\mu=1$. A symmetric cost function implies $C^{C}(2,1)=C^{C}(0,1)$. Yet $\hm=0$ corresponds to perceiving the issue as not complex at all, so any  $s\neq 0$ leads to updating fully to either $0$ or $1$. In contrast, $\hm=2$ reflects only a moderate increase in perceived complexity.}
This implies the cost function is compressed for downwards distortions.  In particular, we assume the following cost function:
\begin{equation}\label{eq:costfunction}
C^{C}(\hm,\mu)=\left\{
\begin{array}{ll}
\frac{c}{2}\left(\hm-\mu\right)^2&\text{if }\hm\geq\mu\\
\frac{c}{2}\left(\frac{\mu}{\hm}\left(\hm-\mu\right)\right)^2&\text{if }\hm<\mu
\end{array}
\right.
\end{equation}
 With this cost function we have  $\left.C^{C}(\hm,\mu)\right|_{\hm=\delta\cdot \mu}=\left.C^{C}(\hm,\mu)\right|_{\hm=\mu/\delta}=\frac{c}{2}\mu^2\left(\delta-1\right)^2$ for all $\delta>1$.
In the left panel of Figure \ref{fig:cost_distortion}, the cost function is plotted for $\mu=1$ and $c=2$. One property of it is that $\lim_{\hm\rightarrow \infty}C^{C}(\hm,\mu)=\lim_{\hm\rightarrow 0}C^{C}(\hm,\mu)=\infty$, and therefore we have $\hm\in(0,\infty)$ for any $s$.

 When deciding how to optimally distort $\mu$, each voter aims to maximize

\[
 W^{CC}=AU\kps-C^{C}(\hm,\mu).
\]
 When  $s=0$, any distortion only causes cost, but does not change $AU\kps$. Hence,  $\hm^*=\mu$.  When signals are maximally informative, the optimal distortion is the same. To see this, suppose $s\rightarrow\infty$. Then \emph{any} distortion except $\hm=0$ will move the belief about $\omega$ to $\pi=1$. But  the cost of choosing $\hm=0$ is infinite, and hence this cannot be optimal. The same is true for $s\rightarrow-\infty$. For any other signal  $s\in\left(-\infty,0\right)\cup\left(0,\infty\right)$, the direction of the optimal distortion is the same as in Lemma \ref{lem:sigma}. 
%Moreover, $ W^C$ is a continuous and smooth function of both $s$ and $\hm$, and hence also $\hm$ is a continuous and smooth function of $s$.
In the right panel of  Figure \ref{fig:cost_distortion}.
 We plotted $\tilde{\mu}^*$ for $\mu=1$ and different values of $c$.

If $s<0$, voters choose optimally at the ballot despite $\hm\neq\mu$. For $s>0$, this is not necessarily the case. These voters interpret signals as less informative than they truly are and may therefore cast a ballot for policy~0 despite $s>0$. Intuitively, as $c$ increases, voters choose to distort $\mu$ less. Consequently, $\lim_{c \to \infty}\hm^*(s)=\mu$. In this case, a majority will always cast a ballot for $p=\omega$, giving candidates an incentive to heed their information. However, if $c$ is small, the outcome reverts to the result in Proposition~\ref{prop:inactivity}. By continuity, there exists a $\hat{c}>0$ such that if $c\in[0,\hat{c}]$, motivated beliefs lead to voting behavior that prevents candidates from following their information. Conversely, if $c>\hat{c}$, then despite choosing $\hm^*\neq\mu$, a majority of voters will vote for policy~1 in state~1, thereby creating incentives for candidates to follow their information and campaign on policy~1.

\subsection{Policy Motivated Politicians}
\label{ssec:policymotiv}

Suppose candidates not only care about being elected but also value welfare, as captured by \eqref{eq:util}. Knowing which state has materialized, candidates are aware of which policy maximizes welfare. Thus, one might hypothesize that adding a policy motivation would improve the situation. 

However, this is only partially correct. On the one hand, if policy motivation is sufficiently important compared to simply being elected to office, there is indeed an equilibrium in which candidates always choose $\mathbf{p}^*=(\omega,\omega)$. To see this, assume the spoils of office have value $V$ and that each candidate receives $V+u(p,\omega)$ if elected and $u(p,\omega)$ otherwise.

Now, suppose candidate 1 expects candidate 2 to choose $p_2=\omega=1$. Choosing $p_1=0$ would win the election but result in a welfare loss due to the wrong policy being chosen. Alternatively, choosing $p_1=1$ reduces the chance of winning the election to 50\%, but it ensures that the optimal policy is selected with certainty. Candidate 1 has a strict incentive to also propose $p_1=1$ if
\[
\frac{V}{2}-\Delta\geq V-\Delta-b \Leftrightarrow b\geq \frac{V}{2}.
\]
If, however, $b<\frac{V}{2}$, there remains a unique \ps equilibrium with $\mathbf{p}^*=(0,0)$. Moreover, increasing the spoils of office $V$ decreases the chance for an efficient equilibrium to exist.

However, even if the spoils of office are sufficiently low, the inefficient equilibrium continues to exist. Assume candidate 1 expects candidate 2 to choose $p_2=0$. In this case, the unique best response is to also choose $p_1=0$. The reason is that now  policy 0 wins with certainty. Candidate 1 cannot influence the policy, but she can increase her chances of winning. Therefore, despite policy motivations, and even in the absence of any office motivation, there always exists an equilibrium with $\mathbf{p}^*=(0,0)$. This  logic is closely related to the one in  \cite{calvert1985robustness}, who shows that the median voter theorem may continue to hold even in the presence of policy motivations.

\section{Conclusion}
\label{sec:conclusion}

This paper illustrates the negative impact that voter beliefs and expectations can have on political decision-making, particularly in the context of climate change policy. The model demonstrates that when voters are inclined to dismiss the severity of climate change, this skews the equilibrium toward suboptimal policy choices by incentivizing office-motivated politicians to act in ways that contradict their own information. The analysis reveals that---in the presence of motivated reasoning---trust in the functioning of government plays a critical role for  the efficiency of policy-making.  Moreover, the paper suggests that  political rhetoric plays an important role as well: if politicians, scientists, journalists, and activists aim to contribute to better climate change policy, they must navigate a delicate balance between conveying information truthfully and avoiding the risk of pushing voters into denial.

\appendix

\section{Mathematical Appendix}
\label{sec:app:proofs}

\subsection{Proof of Lemma \ref{lem:sigma}}
The proof follows from \eqref{eq:dAUdsigma}:
\[
\begin{array}{rcl}
\dfrac{\partial  AU(s,\mathbf{p},\tilde{\mu})}{\partial \hm}>0
\Leftrightarrow -\dfrac{2 (1-q) q (1-\rho_0) \rho_0 (1-\rho_1) \rho_1 s \left[b +\Delta -\kappa(1+b) \right] e^{\frac{2 \hm  s}{\sigma ^2}}}{\sigma^2\left(q\rho_1(1-\rho_1)+(1-q)\rho_0(1-\rho_0)e^{-\frac{2 \hm  s}{\sigma^2}}\right)^2}&>&0\\
%\Leftrightarrow -s \left[b +\Delta -\kappa(1+b) \right]&>&0\\
\Leftrightarrow s \left[b +\Delta -\kappa(1+b) \right]&<&0
\end{array}
\]
The realization of $s$ and the expression in brackets  jointly determine the derivative's sign. Consider next the term in parentheses:
\[
%\begin{array}{rcl}
b +\Delta -\kappa(1+b) >0
\Leftrightarrow \kappa<\tilde{\kappa}\equiv\frac{\Delta+b}{1+b}
%\end{array}
\]

If $\Delta>1$, then $\tilde{\kappa}>1$, and hence we must have  $\kappa<\tilde{\kappa}$. It follows that   $\text{Sign}\left[\dfrac{\partial  AU(s,\mathbf{p},\tilde{\mu})}{\partial \hm}\right]=-\text{Sign}\left[s\right]$. Hence, $\hm^*=\infty$ if $s<0$ and $\hm^*=0$ if $s>0$. If $s=0$, any $\hm$ yields the same $AU$ and hence  $\hm^*=\sigma$.

If $\Delta<1$, then $\tilde{\kappa}<1$ as well. Hence, if $\kappa<\tilde{\kappa}$, then again
$\text{Sign}\left[\dfrac{\partial  AU(s,\mathbf{p},\tilde{\mu})}{\partial \hm}\right]=-\text{Sign}\left[s\right]$. Thus,  $\hm^*=\infty$ if $s<0$,  $\hm^*=0$ if $s>0$, and $\hm^*=\sigma$ if  $s=0$. To the contrary, if  $\kappa>\tilde{\kappa}$, then
$\text{Sign}\left[\dfrac{\partial  AU(s,\mathbf{p},\tilde{\mu})}{\partial \hm}\right]=\text{Sign}\left[s\right]$. Hence,   $\hm^*=0$ if $s<0$,  $\hm^*=\infty$ if $s>0$, and $\hm^*=\sigma$ if  $s=0$. If $\kappa=\tilde{\kappa}$, then $\text{Sign}\left[\dfrac{\partial  AU(s,\mathbf{p},\tilde{\mu})}{\partial \hm}\right]=0$, and any $\hm$ yields the same anticipatory utility. Thus,  $\hm^*=\mu$.

Finally, if $\Delta=1$, then $\tilde{\kappa}=1$, and thus we cannot have $\kappa>\tilde{\kappa}$. If $\kappa=1$, then  $AU\kps$ is  flat in $\hm$, implying any $\hm^*=\mu$. If $\kappa<1$, then the situation resembles the case of $\Delta>1$.\qed

\subsection{Proof of Corollary \ref{cor:beliefs}}
Follows from the discussion in the text and Lemma \ref{lem:sigma}.\qed

\subsection{Proof of Proposition \ref{prop:voting}}

 We prove the different cases successively.

\paragraph{$\Delta>1$:}  We know from Corollary  \ref{cor:beliefs} that all voters with $s>0$  have belief $\pi\kps=\hat{\pi}$, while those with $s<0$  have belief  $\pi\kps=0$.
Hence, the latter always vote for policy 0. whether a voter with $s>0$ votes for policy 1 or policy 0 depends on the comparison of $\tp$ and $\hp$.  In particular, we know from \eqref{eq:voting_rule} that a voter votes for policy 1 if $\pi\kps>\tp$, for policy 0 if $\pi\kps<\tp$, and they choose each with a probability of 50\% if $\pi\kps=\tp$. Hence, if $\hp<\tp$, all voters vote for policy 0. If $\hp=\tp$, all voters receiving $s<0$ plus half of the voter receiving $s>0$ vote for policy 0, while the other half of $s>0$ voters vote for policy 1. As a consequence, policy 0 wins all the time if $\hp\leq \tp$.
 This implies that we must have $\kappa^*=0$ for all $s$.

However, if $\hp>\tp$, all voters vote informatively, and thus the optimal policy always wins.
It follows that in state 0 a majority votes for policy 0, while in state 1 a majority votes for policy 1. Hence, voters know that the policy that matches the state wins all the time. The belief about the state is $\pi\kps=\hp$ if $s\geq0$, implying $\left.\kappa^*\right|_{s\geq0}=\hp$, and $\pi\kps=0$ if $s<0$, and hence $\left.\kappa^*\right|_{s<0}=0$.

Could there be other pure strategy equilibria?
 $\hm^*$ is uniquely determined, and hence also $\pi\kpss$ is uniquely determined. But this implies that equilibrium vote shares are  unique, leaving no room for beliefs $\kappa$ that are consistent with these vote shares and that differ from the ones established before. Hence, if $\Delta>1$, no other equilibrium in the voting subgame can exist.

\paragraph{$\Delta=1$:}
If $\kappa=1$, then $\pi\kps=\pi(s,\mathbf{p},\mu)$. 
Hence, any voter with $s<0$ will always vote for policy 0. This means policy 0 wins at least if $\omega=0$, contradicting $\kappa=1$. Hence, if $\Delta=1$, there cannot be an equilibrium with $\kappa=1$.

If $\kappa<1=\tilde{\kappa}$, then $\hm^*=0$ if $s>0$ and $\hm^*=\infty$ if $s<0$. It follows that equilibrium is the same as in the case of $\Delta>1$.

\paragraph{$\Delta< 1$:}

%
% If $\Delta\leq 1$, then there are there exist multiple subgame equilibria.
%  \begin{enumerate}
%    \item For any $\hat{\pi}\in(0,1)$, there exists a subgame equilibrium  with  $\kappa^*\kps =0$ if $s< 0$ and $\kappa^*\kps =1$ if $s\geq 0$, and a majority of voters always votes for the policy $p_i$  that matches the true state $\omega$.
%    \item For $\hat{\pi}\geq\tilde{\pi}$, there exists a subgame equilibrium with $\kappas=1$ and policy 1 always wins.
%    \item For $\hat{\pi}\leq\tilde{\pi}$, there exists a subgame equilibrium with $\kappas=0$ and policy 0 always wins.
%    \item There is no subgame equilibrium in which the  policy $p_i$ that does not match the state $\omega$ always  wins.
%  \end{enumerate}
First assume that $\kappa$ positively correlates with $s$ in the sense that $\text{Sign}\left[\kappa-\tilde{\kappa}\right]=\text{Sign}\left[s\right]$. Any voter with $s<0$ chooses $\hm^*=\infty$, and thus all these voters hold belief $\pi\kpss=0$ and vote for policy 0. Moreover, any voter with $s>0$ chooses also $\hm^*=\infty$, and thus all these voters hold belief $\pi\kpss=1$ and vote for policy 1. This implies that indeed the policy that matches the state always wins and we need to have $\left.\kappa\right|_{s<0}=0$ as well as $\left.\kappa\right|_{s>0}=1$, proving part 2 (a) of the proposition.

Next, assume voters generally believe that policy 0 will be implemented, independent of the true state $\omega$. Then $\kappa=0$ for all $s$. It follows that any voter with $s<0$ chooses $\hm^*=\infty$, and thus all these voters hold belief $\pi\kpss=0$ and vote for policy 0. Moreover, any voter with $s>0$ chooses  $\hm^*=0$, implying $\pi\kpss=\hp$. Iff $\hp\leq \tp$, this implies that indeed always a majority of voters votes for policy 0, and hence when $\hp\leq \tp$ there exists an equilibrium of the voting subgame in which policy 0 is always chosen. This proves part 2 (b).

Now assume voters generally believe that policy 1 will be implemented, independent of the true state $\omega$. Then $\kappa=1$ for all $s$, and  any voter with $s>0$ chooses $\hm^*=\infty$, and thus all these voters hold belief $\pi\kpss=1$ and vote for policy 1. Moreover, any voter with $s<0$ chooses  $\hm^*=0$, inducing $\pi\kpss=\hp$. If and only if $\hp\geq \tp$, this implies that indeed always a majority of voters votes for policy 1, and hence when $\hp\geq \tp$ there exists an equilibrium of the voting subgame in which policy 1 is always chosen. This proves part 2 (c).

Finally, consider part 2 (d).
An equilibrium in which the optimal policy never wins would imply  policy 1 wins if $\omega=0$ and policy 0 wins if $\omega=1$. Hence, we would need to have $\kappa$ weakly decreasing in $s$. Consider a voter with signal $s>0$. It must be true that most of these voters vote for policy 0. Hence, it cannot be true that $\kappa\geq \tilde{\kappa}$. If $\kappa<\tilde{\kappa}$, these voters all hold belief $\hp$. Next consider voters with $s<0$. It must be true that most of these voters vote for policy 1, implying $\kappa\leq \tilde{\kappa}$ is not possible.  Thus, because $\kappa>\tilde{\kappa}$, these voters all also hold belief $\hp$. If $\hp>\tp$, policy 1 always wins, contradicting that the optimal policy never wins.
If $\hp<\tp$, policy 0 always wins, also contradicting that the optimal policy never wins. Finally, if $\hp=\tp$, in each state, each policy wins with 50\%, and thus also  the optimal  policy wins with a chance of 50\%. This implies that the actual state is irrelevant for the probability of each policy winning, and thus we must have $\kappa=\ot$ for all $s$. But then it is not possible that $\kappa$ differs for voters with $s>0$ and voters with $s<0$. Hence, such an equilibrium of the voting subgame cannot exist.

\medskip

This completes the proof of the proposition.\qed

\subsection{Proof of Proposition \ref{prop:cands}}
Follows from the discussion in the text.\qed

\subsection{Proof of Proposition \ref{prop:inactivity}}
In a symmetric equilibrium, each candidate wins with a 50\% probability.  We next show that no lucrative deviations exist. Hence, suppose that off the equilibrium path voters observe $\tilde{\mathbf{p}}\in\{(0,1),(1,0)\}$.

\paragraph{$\Delta\geq1$:}

 We know from Corollary \ref{cor:beliefs} that after  $\pi(s,\tilde{\mathbf{p}},\hm^*)=0$ if $s<0$ and $\pi(s,\tilde{\mathbf{p}},\hm^*)=\hp$ if $s\geq0$. Hence,  $\hp\leq \tp$, and thus no voter ever votes for policy 1. It follows that policy 1 never wins. Consequently, a deviation from $\mathbf{p}=(0,0)$ is never profitable, and thus $\mathbf{p}^*=(0,0)$ is indeed an equilibrium. Moreover, $(1,1)$ cannot be an equilibrium, because a deviation is always profitable, because when spotting such a deviation, voters will ignore information implying that climate change is severe.
Thus, the unique symmetric equilibrium is candidates choosing   $\mathbf{p}^*=(0,0)$.  This is supported by $\kappa^*=0$ and $\hm^*(s)$ as described in Lemma \ref{lem:sigma}.

\paragraph{$\Delta< 1$:}
%It is straightforward to show that the equilibrium described above for the case $\Delta>1$ still exists.
It is easy to show that the inefficient equilibrium with $\mathbf{p}^*=(0,0)$ still exists. Hence, here we focus on the existence of the efficient equilibrium.

%Assume first $\kappa=0$ for all $s$. Then it follows from Corollary \ref{cor:beliefs} that $\pi(s,\tilde{\mathbf{p}},\hm)=0$ if $s<0$ and $\pi(s,\tilde{\mathbf{p}},\hm)=\hp=q<\tp$ if $s>0$. Thus, a majority always votes for policy 0. It follows from Proposition \ref{prop:cands} that no candidate has an incentive to campaign on $p=1$. Consequently, $\mathbf{p}^*=(0,0)$, $\kappa^*=0$, and $\hm^*$ as described in Lemma \ref{lem:sigma} is an equilibrium.

Suppose candidates choose $\mathbf{p}=(1,1)$. Hence, as before, each wins with a 50\% probability. If a candidate deviates, they face conflicting policy platforms, and hence do not learn from $\mathbf{p}$ itself. The only difference between voters is the signals they receive. Hence, if they hold different beliefs, these need to be based on signals.

Suppose voters expect the winning candidate to be the candidate offering $p=\omega$. That means 
voters have $\kappa=\pi(s)$. Suppose 
that $\kappa>\tilde{\kappa}$ if $s>0$ and $\kappa<\tilde{\kappa}$ if $s<0$. Then it follows from Corollary \ref{cor:beliefs} that $\pi(s,\tilde{\mathbf{p}},\hm)=0$ if $s<0$ and $\pi(s,\tilde{\mathbf{p}},\hm)=1$ if $s>0$. Hence, all voters vote informatively, and thus the policy that matches the state wins. This in turn implies that indeed  $\kappa=\pi(s,\tilde{\mathbf{p}},\hm)$.  It follows from Proposition \ref{prop:cands} that, if candidates expect voters to behave this way, they have an incentive to choose the socially optimal policies. Hence, $\mathbf{p}^*=(\omega,\omega)$, $\kappa^*=\pi(s,\mathbf{p}^*,\hm^*)\in\{0,1\}$ for all $s$,  and $\hm^*=\infty$ for all $s\neq 0$ is also an equilibrium if $\Delta<1$.

It follows from Proposition \ref{prop:voting} part 2 (d) that no equilibrium can exist in which always the wrong policy is chosen. Hence, we now only need to show that no equilibrium, in which policy 1 always wins, can exist.
 Suppose to the contrary that voters expect policy 1 to be implemented   with certainty, $\kappa=1\geq\tilde{\kappa}$ for all $s$. Then it follows from Corollary \ref{cor:beliefs} that $\pi\kps=q< \tp$ if $s<0$ and $\pi\kps=1$ if $s>0$. Therefore,  all voters vote informatively, implying $p=\omega$ wins if $p_1\neq p_2$, thus contradicting $\kappa=1$.
  Hence, it follows that no equilibrium, in which policy 1 is always chosen, exists. 
  
  This proves the proposition.
\qed
%
%
%\subsection{Proof of Lemma \ref{lem:fake_alpha}}
%Take the belief about $\omega$ of a voter who receives $s$ and chooses $\tilde{\alpha}$:
%\[
%\pi(s,\tilde{\alpha})=\frac{q\left(\tilde{\alpha} g(s)+(1-\tilde{\alpha})f_1(s)\right)}{q\left(\tilde{\alpha} g(s)+(1-\tilde{\alpha})f_1(s)\right)+(1-q)\left(\tilde{\alpha} g(s)+(1-\tilde{\alpha})f_0(s)\right)}.
%\]
%Anticipatory utility is defined as before in \eqref{EQ:AU}. Hence, it's derivative with respect to  $\tilde{\alpha}$
%is
%\[
%\dfrac{\partial AU}{\partial \tilde{\alpha}}=\frac{(1-q) q g(s) \left[b +\Delta-(b +1) \kappa\right] (f_1(s)-f_0(s))}{\left[q\left(\tilde{\alpha} g(s)+(1-\tilde{\alpha})f_1(s)\right)+(1-q)\left(\tilde{\alpha} g(s)+(1-\tilde{\alpha})f_0(s)\right)\right]^2}
%\]
%The derivative is positive if $f_1(s)-f_0(s)>0\Leftrightarrow s>0$, negative if   $f_1(s)-f_0(s)<0\Leftrightarrow s<0$, and zero else. This proves the result. \qed
%
%
%%
%%
%%\subsection{Proof of Proposition \ref{prop:fake}}
%%It follows from Lemma \ref{lem:fake_alpha} that independent of the true state $\omega$, a majority of voters prefers to vote for policy 0 when offered the choice. The proof then follows from Proposition \ref{prop:cands}.
%%\qed

\section{Costly Distortions}
\label{app:costly}
\subsection{Cost from Imperfect Decision Making}
Recall that voters aim at maximizing $ W^{UC}=\lambda AU+(1-\lambda)C^U$. Each voter chooses $\hm$ to maximize $ W$, taking into account that $\hm\neq \mu$ may negatively impact the quality of the decision at the ballot.

 When $\hm$ is chosen in a way that is inconsequential for the decision at the ballot, $C^U=0$. Otherwise, $C^U$ equals the expected cost from deciding incorrectly.

If a voters chooses to vote for policy 1 at the ballot, her expected utility is 
\[
u_1=\frac{-\Delta  q e^{\frac{2 \mu  s}{\sigma ^2}}+q-1}{q \left(e^{\frac{2 \mu  s}{\sigma ^2}}-1\right)+1}.
\]
 When voting for policy 0, expected utility is
\[
u_0=-\frac{q (b +\Delta )}{q-(q-1) e^{-\frac{2 \mu  s}{\sigma ^2}}}
\]
Hence, a voter votes for policy 1 iff
\[
u_1\geq u_0\Leftrightarrow s>\tilde{s}:=\frac{\sigma ^2 \log \left(\frac{1-q}{b  q}\right)}{2 \mu }>0.
\]
Any voter with $s<\tilde{s}$ will vote for policy 0 with Bayesian information processing, and this does not change with motivated beliefs. Thus, If $s<\tilde{s}$, then $C^U=0$ because the choice at the ballot is identical to the Bayesian optimal choice. If $s>\tilde{s}$, this may change, as the Bayesian optimal choice would be to vote for policy 1, whereas with distorted information processing policy 0 might be chosen. 
If the choice at the ballot is indeed altered by choosing $\hm\neq \mu$, then the voter incurs a cost of 
\[
C^U=-b +\frac{(b +1) (1-q)}{1-q+q e^{\frac{2 \mu  s}{\sigma ^2}}}.
\]
This equals the expected utility loss from taking an incorrect decision.

As explained in the main text, when deciding if and how to distort beliefs, a voter has only two  potentially optimal choices:  $\hm=0$ and $\hm=\mu_c$, where $\mu_c$
is the distortion that induces a posterior belief of $1/(1+b)$. These lead to the following welfare levels:
\[
\begin{array}{rcl}
 W^{UC}(\hm=0)&=&-(1-\lambda) \left(b +\frac{(b +1) (q-1)}{q \left(e^{\frac{2 \mu  s}{\sigma ^2}}-1\right)+1}\right)-\lambda  q (b +\Delta )\\
 W^{UC}(\hm=\mu_c)&=&-\frac{\lambda  (b +\Delta )}{b +1}
\end{array}
\]
Note that 
\[
\left. W^{UC}(\hm=0)\right|_{s=\tilde{s}} -  W(\hm=\mu_c) = \frac{\lambda(b+\Delta)(1-b q - q)}{b + 1} > 0
\]
for all $\lambda>0$. Moreover,
\[
\frac{\partial  W^{UC}(\hm=0)}{\partial s} = -\frac{2(b + 1)(1 - \lambda)\mu(1 - q)q e^{\frac{2\mu s}{\sigma^2}}}{\sigma^2 \left[q \left(e^{\frac{2\mu s}{\sigma^2}} - 1\right) + 1\right]^2} < 0.
\]
Therefore, if 
\[
\begin{array}{c}
\lim_{s \to \infty}  W^{UC}(\hm=0) >  W(\hm=\mu_c) 
\Leftrightarrow b(\lambda + \lambda(-q) - 1) - \Delta\lambda q > -\frac{\lambda(b+\Delta)}{b + 1} \\
\Leftrightarrow \lambda > \tilde{\lambda} := -\dfrac{b(b + 1)}{\Delta(b q + q - 1) + b(b(q - 1) + q - 2)} \in (0,1),
\end{array}
\]
then any voter with $s>0$ will choose $\hm^* = 0$. Conversely, if $\lambda < \tilde{\lambda}$, there exists $\hat{s} > \tilde{s}$, where
\[
\hat{s} = \frac{\sigma^2}{2\mu} \log\!\left(\frac{(q-1)\!\left[-\Delta\lambda + \lambda + b^2 \lambda q + b\big((\Delta + 1)\lambda q - 1\big) + \Delta\lambda q - 1\right]}{q\!\left[b + b^2(\lambda(q - 1) + 1) + b\lambda(\Delta q + q - 2) + \Delta\lambda(q - 1)\right]}\right),
\]
such that $\hm^* = 0$ if $s \in [0,\hat{s}]$ and $\hm^* = \mu_c$ if $s > \hat{s}$.

If $\lambda>\tilde{\lambda}$, then all results are as in the baseline version of the model. Otherwise, whether or not candidates will ignore their information depends on the complexity of climate change. 
For candidates to have incentives to offer policy~1, at least 50\% of voters must receive $s > \hat{s}$ if $\omega=1$. If $\mu$ is sufficiently large, this condition is satisfied. However, if complexity is high, this will not hold, and the result from the baseline model continues to apply.

\subsection{Cognitive Costs}
Following the discussion in the main text, we now establish a useful lemma:
\begin{lemma}
\label{lem:costly2}
Suppose $\Delta>1$. If $s\in\{-\infty,0,\infty\}$, then $\hm(s)=\mu$. Otherwise, and if $s>0$, then $\hm(s)<\mu$, and there exists $\bar{s}>0$ such that $\hm(s)$ decreases in $s$ if $s\in[0,s^+)$, and it increases in $s$ if $s>s^+$. If $s<0$, then $\hm(s)>\mu$, and there exists $s^-<0$ such that $\hm(s)$ decreases in $s$ if $s\in(s^-,0]$, and it increases in $s$ if $s<s^-$. Moreover, for every $s$, the absolute distortion $\left|\mu-\hm\right|$ decreases in $c$
\end{lemma}

\begin{proof}
To see incentives, first consider how $ W^{CC}$ changes with $\hm$, when $\hm\geq\mu$. It is easy to show that $ W^{CC}$ is strictly concave in $\hm$ and thus an interior equilibrium exists for all $s$. The FOC for an interior optimum is
\[
\left.\frac{\partial  W^{CC}\kps}{\partial \hm}\right|_{\hm\geq\mu}=-\frac{2 (1-q) q s \left[b +\Delta-(b +1) \kappa  \right] e^{\frac{2
{\hm} s}{\sigma ^2}}}{\left(q \sigma  \left(e^{\frac{2 \hm s}{\sigma ^2}}-1\right)+\sigma \right)^2}-c (\hm-\mu).
\]
if $\hm>\mu$ and
\[
\left.\frac{\partial  W^{CC}\kps}{\partial \hm}\right|_{\hm<\mu}=-\frac{2 (1-q) q s \left[b +\Delta-(b +1) \kappa  \right] e^{\frac{2
{\hm} s}{\sigma ^2}}}{\left(q \sigma  \left(e^{\frac{2 \hm s}{\sigma ^2}}-1\right)+\sigma \right)^2}-c\frac{\mu^3}{\hm^3} (\hm-\mu).
\]
if $\hm<\mu$.
In both derivatives, the first term's sign  is the opposite of the sign of $s$ (when $\Delta>1$). Hence, if $s>0$, then $\hm^*(s)<\mu$. Similarly, if $s<0$, we must have $\hm^*(s)>\mu$. If $s=0$, the first term is zero, and hence we need to have $\hm^*(s)=\mu$.
Moreover, $
\lim_{s\rightarrow\infty}\left.\frac{\partial  W^{CC}\kps}{\partial \hm}\right|_{\hm<\mu}=-c\frac{\mu^3}{\hm^3}(\hm-\mu)$,
and hence  $\lim_{s\rightarrow\infty}\hm^*(s)=\mu$. Similarly, $\lim_{s\rightarrow-\infty}\left.\frac{\partial  W^{CC}\kps}{\partial \hm}\right|_{\hm\geq\mu}=-c(\hm-\mu)$,
and hence  also $\lim_{s\rightarrow-\infty}\hm^*(s)=\mu$.

Now consider $s<0$, for which we have $\hm>\mu$. Rearranging the FOC yields
\[
-\frac{2 (1-q) q s \left[b +\Delta-(b +1) \kappa  \right] e^{\frac{2
{\hm} s}{\sigma ^2}}}{\left(q \sigma  \left(e^{\frac{2 \hm s}{\sigma ^2}}-1\right)+\sigma \right)^2c}+\mu=\hm
\]
That is, we can express $\hm$ as a deviation from $\mu$, and
\[
\Gamma=-\frac{2 (1-q) q s \left[b +\Delta-(b +1) \kappa  \right] e^{\frac{2
{\hm} s}{\sigma ^2}}}{\left(q \sigma  \left(e^{\frac{2 \hm s}{\sigma ^2}}-1\right)+\sigma \right)^2c}
\]
defines the deviation. $\Gamma$ increases in $s$ iff
\[
\frac{\partial \Gamma}{\partial s}=\frac{-2 (1-q) q \left[b +\Delta-(b +1) \kappa\right] e^{\frac{2 \hm s}{\sigma ^2}} \left((1-q) \left(\sigma ^2+2 \hm s\right) -qe^{\frac{2 \hm s}{\sigma ^2}} \left(2 \hm s-\sigma ^2\right)\right)}{c \sigma ^4 \left(q
   \left(e^{\frac{2 \hm s}{\sigma ^2}}-1\right)+1\right)^3}>0,
\]
which  is the case iff $
\Omega=(1-q) \left(\sigma ^2+2 \hm s\right) -qe^{\frac{2 \hm s}{\sigma ^2}} \left(2 \hm s-\sigma ^2\right)<0$.
If $s=0$, $\left.\Omega\right|_{s=0}=\sigma ^2>0$, and hence $\hm^*(s)$ is decreasing. Moreover, $\lim_{s\rightarrow-\infty}\Omega=-\infty$. Thus, if $\Omega$ is monotone in $s$, then there is a unique $s^-<0$ such that $\hm^*(s)$ increases in $s$ if $s<s^-$ and it decreases in $s$ if $s\in(s^-,0]$.  We have
\[
\frac{\partial \Omega}{\partial s}=2 \hm \left(1-q-q\frac{2 \hm s e^{\frac{2 \hm s}{\sigma ^2}}}{\sigma ^2}\right)>0.
\]
This proves the existence of a unique $s^-$. Moreover, because $\left|\Gamma\right|$ decreases in $c$, the absolute deviation from $\mu$ decreases in the cost $c$.

Next consider $s>0$, implying $\hm<\mu$. Rearranging the FOC yields
\[
-\frac{2 \hm^3 (1-q) q s \left[b +\Delta -(b +1) \kappa\right] e^{\frac{2 \hm s}{\sigma ^2}}}{c \mu ^3 \left(q \sigma  \left(e^{\frac{2 \hm s}{\sigma ^2}}-1\right)+\sigma \right)^2}+\mu =\hm
\]
That is, we can again express $\hm$ again as a deviation from $\mu$, and
\[
\Gamma'=-\frac{2 \hm^3 (1-q) q s \left[b +\Delta -(b +1) \kappa\right] e^{\frac{2 \hm s}{\sigma ^2}}}{c \mu ^3 \left(q \sigma  \left(e^{\frac{2 \hm s}{\sigma ^2}}-1\right)+\sigma \right)^2}
\]
defines the deviation. $\Gamma'$ increases in $s$ iff
\[
\frac{\partial \Gamma'}{\partial s}=
\frac{2 \hm^3 (1-q) q  \left[b +\Delta -(b +1) \kappa\right] e^{\frac{2 \hm s}{\sigma ^2}} \left(q e^{\frac{2 \hm s}{\sigma ^2}} \left(2 \hm s-\sigma ^2\right)-(1-q) \left(\sigma ^2+2 \hm s\right)\right)}{c \sigma ^4 \left(\mu +\mu  q
   \left(e^{\frac{2 \hm s}{\sigma ^2}}-1\right)\right)^3}>0
\]
This is the case iff $\Omega'=q e^{\frac{2 \hm s}{\sigma ^2}} \left(2 \hm s-\sigma ^2\right)-(1-q) \left(\sigma ^2+2 \hm s\right)>0$.
If $s=0$, $\left.\Omega\right|_{s=0}=-\sigma ^2<0$, and hence $\hm^*(s)$ is decreasing. Moreover, $\lim_{s\rightarrow\infty}\Omega'=\infty$. Finally,
\[
\frac{\partial \Omega'}{\partial s}=\frac{2 \hm q s e^{\frac{2 \hm s}{\sigma ^2}}}{\sigma ^2}+q-1
\]
This is negative when $s=0$. Hence, if we can show that $\Omega'$ is convex, we  prove the existence of a unique $s^+$. Take the second derivative with respect to $s$:
\[
\frac{\partial^2 \Omega'}{\partial s^2}=\frac{2 \hm q e^{\frac{2 \hm s}{\sigma ^2}} \left(\sigma ^2+2 \hm s\right)}{\sigma ^4}>0,
\]
and thus $\Omega'$ is indeed convex in $s$. Thus, there exists a unique $s^+>0$ such that $\hm^*(s)$ increases in $s$ if $s>s^+$ and it decreases in $s$ when $s\in[0,s^+)$. Moreover, because $\left|\Gamma'\right|$ decreases in $c$, the absolute deviation from $\mu$ decreases in the cost $c$. This proves the lemma.
\end{proof}
The remaining discussion from the main text follows from the lemma.

\end{document}